\documentclass[sigconf]{acmart}
\copyrightyear{2026}
\acmYear{2026}
\setcopyright{cc}
\setcctype{by}
\acmConference[DAC '26]{63rd ACM/IEEE Design Automation Conference}{July 26--29, 2026}{Long Beach, CA, USA}
\acmBooktitle{63rd ACM/IEEE Design Automation Conference (DAC '26), July 26--29, 2026, Long Beach, CA, USA}
\acmDOI{10.1145/3770743.3812057}
\acmISBN{979-8-4007-2254-7/2026/07}

\title[LLM for EDA in Front-End Design: Challenges and Opportunities]
{LLM for EDA in Front-End Design: Challenges and Opportunities\vspace{-0.2cm}}

\usepackage{threeparttable}
\usepackage{array}
\usepackage{subfigure}
\usepackage{tablefootnote}
\usepackage{booktabs}
\usepackage{makecell}
\usepackage{stfloats}
\usepackage{graphicx}
\usepackage{textcomp}
\usepackage{xcolor}
\usepackage{pifont}
\usepackage{algorithm}
\usepackage[normalem]{ulem}
\usepackage{algpseudocode}
\usepackage{multirow}
\usepackage{fix-cm}

\begin{document}

\author{\vspace{0cm}Kangwei Xu\textsuperscript{1}, Bing Li\textsuperscript{2}, Ulf Schlichtmann\textsuperscript{1}\\}
\affiliation{%
  \textsuperscript{1} Chair of Electronic Design Automation, \textit{Technical University of Munich (TUM)}, Munich\country{Germany}\\
  \textsuperscript{2} Resource-Efficient AI Group, \textit{Technical University of Ilmenau}, Ilmenau\country{Germany}\\
  \mbox{Email:\hspace{0.3em}}\{kangwei.xu, ulf.schlichtmann\}@tum.de, bing.li@tu-ilmenau.de}

\renewcommand{\shortauthors}{Kangwei Xu, Bing Li, Ulf Schlichtmann}
\begin{abstract} 
As chip complexity increases and time-to-market pressures grow, front-end design has become a critical bottleneck in chip development. 
Recently, Large Language Models (LLMs) have shown great potential in Electronic Design Automation (EDA). Beyond specification understanding, LLMs show the potential to serve as a unified intelligent interface for hardware description language (HDL) generation, testbench construction, and design space exploration.
The rise of agentic AI, represented by pioneering systems such as OpenClaw, offers a strategic roadmap for the next generation EDA. From this perspective, this paper discusses the evolution of EDA from localized assistance to autonomous agentic execution. 
Then, we review representative advances of LLMs in front-end design, focusing on key tasks such as circuit and testbench generation from a shared specification, as well as design quality improvement in established workflows such as high-level synthesis.
Finally, we discuss the key challenges and limitations of integrating LLMs into EDA, and outline future opportunities for advancing LLM-enabled front-end design, offering a systematic perspective for researchers interested in leveraging agentic AI technologies for EDA.

\end{abstract}

\maketitle

\vspace{-0.2cm}
\begingroup
\small
\noindent\textbf{ACM Reference Format:}\\
Kangwei Xu, Bing Li, and Ulf Schlichtmann. 2026. LLM for EDA in Front-End Design: Challenges and Opportunities.
In \textit{63rd ACM/IEEE Design Automation Conference (DAC '26), July 26 -- July 29, 2026,
Long Beach, CA, USA}. ACM, New York, NY, USA, 5 pages.
\url{https://doi.org/10.1145/3770743.3812057}
\par
\endgroup

\vspace{-0.2cm}
\section{Introduction}
The rapid evolution of Large Language Models (LLMs) is propelling Electronic Design Automation (EDA) into a new technological frontier. Across the full chip design flow, front-end design is especially well aligned with LLM capabilities, because it relies heavily on natural language understanding and high-level logical reasoning. By learning from large collections of hardware description language (HDL) designs, LLMs can capture useful semantics and practical knowledge. As a result, the next generation of EDA is gradually moving beyond the traditional script-driven paradigm toward a more intelligent and automated workflow.

\begin{figure}[t]
\vspace{0.1cm}
\centering	\includegraphics[width=1.03\linewidth]{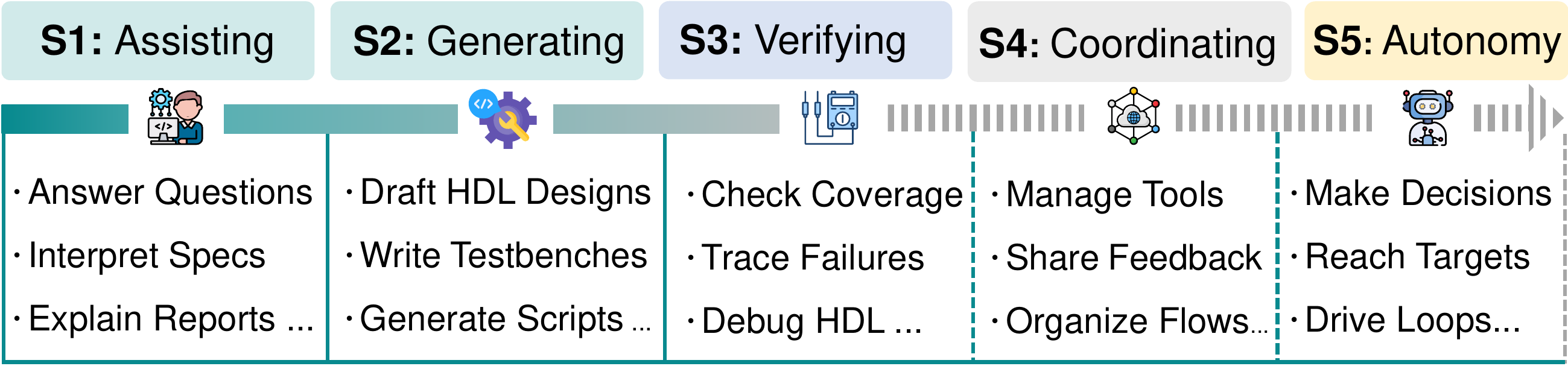}
	\vspace{-0.65cm}
	\caption{~Evolution Path of LLM Capabilities in Chip Design.}
	\label{fig:path}
    \vspace{-0.2cm}
\end{figure}

Recent studies have demonstrated promising results across a wide range of front-end design tasks. As illustrated in Fig.~\ref{fig:path}, these tasks range from assistance-oriented such as question answering, specification interpretation \cite{8.4}, and report explanation \cite{8.70}, to more generation-oriented tasks, including HDL generation \cite{4}, testbench construction \cite{13.1,14}, and script development \cite{16.00}. Initial progress has also been demonstrated in HDL debugging tasks \cite{16.1,16.2}. These results indicate that LLMs are already effective for localized front-end tasks that involve high-level reasoning and explicit feedback.

Despite this progress, moving beyond early-stage assistance and generation remains difficult. The core challenge is not merely the automation of individual tasks, but preserving semantic consistency across different design stages. In hardware design, an implementation may be syntactically correct and pass basic test vectors, while still containing errors in timing boundaries, concurrent interactions, or corner cases. More importantly, even minor semantic mismatches introduced in early stages can propagate through downstream design steps and become harder to detect and debug later. By the time they are exposed during downstream verification, they no longer appear in their original high-level form, but only as low-level failure symptoms, making it difficult to trace them back to where the semantic deviation was first introduced.

\begin{figure}[t]
\centering	\includegraphics[width=1.03\linewidth]{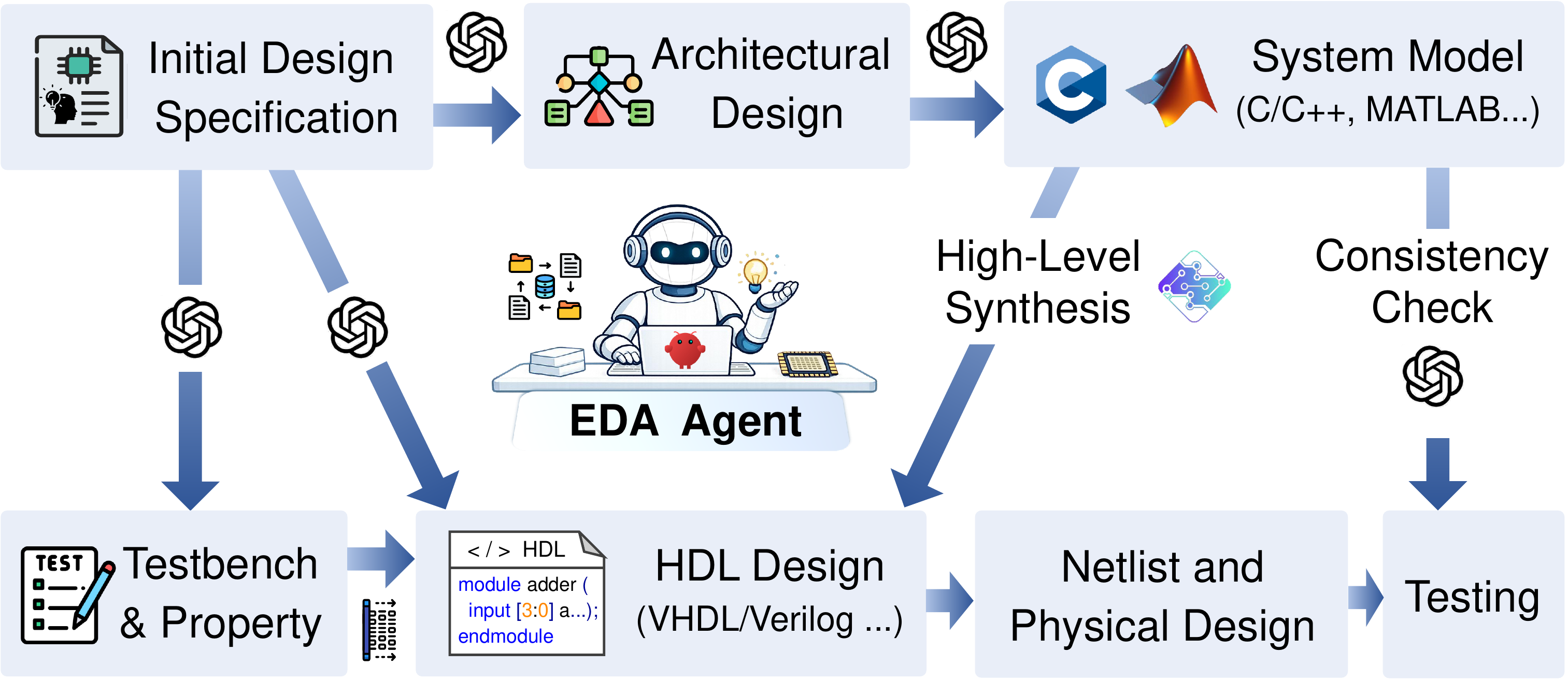}
	\vspace{-0.65cm}
	\caption{~LLM-Based Agentic Chip Design Flow.}
	\label{fig:agentic}
    \vspace{-0.55cm}
\end{figure}

The rise of agentic AI provides a new system-level perspective for addressing this problem. Representative systems such as OpenClaw \cite{18.3} demonstrate that the value of an intelligent agent lies not only in generating text, but also in orchestrating tool usage, maintaining long-horizon task consistency, leveraging persistent memory, and continuously advancing complex workflows. These properties are particularly important in EDA, where design, verification, and optimization are tightly coupled and inherently iterative.

From this perspective, the EDA system requires a robust closed-loop mechanism that can support the full process from design generation to automated debugging. Such a system could work in a way closer to an experienced engineer: decomposing complex requirements into manageable subtasks, internalizing HDL design heuristics, and iteratively debugging and refining designs based on feedback from external EDA tools. With accumulated design knowledge and retrievable past experience, LLMs may evolve from local assistants into autonomous agents capable of driving the front-end design flow, while human engineers focus increasingly on high-level decisions and design objectives.

Motivated by this trend, this paper provides a systematic perspective on LLM applications in front-end design, with a particular focus on whether these advances represent a genuine technological breakthrough or primarily reflect optimistic expectations associated with the rise of agentic AI technologies. 

The rest of this paper is organized as follows. Section~\ref{sec:1second} reviews representative progress in LLM-based front-end design workflows. Section~\ref{sec:1third} presents case studies on HDL generation and testing, and further extends the discussion to High-Level Synthesis (HLS). Section~\ref{sec:fourth} discusses future opportunities and key challenges for LLMs in EDA. Section~\ref{sec:fifth} concludes the paper.

\vspace{-0.4cm}
\section{State of the Art of LLM Applications in EDA} \label{sec:1second}
Chip design is a staged refinement process that progressively transforms natural-language specifications into precise hardware implementations, while maintaining semantic consistency and enabling verification convergence throughout the flow.

As illustrated in Fig.~\ref{fig:agentic}, the process begins with the design specification, which is typically captured in text documents, presentation slides, or design notes. These artifacts describe functional objectives, performance targets, and design assumptions. At this stage, LLMs are primarily used to interpret design intent, and clarify descriptions into more structured inputs \cite{8.4}.

The next stage is architectural design, where high-level requirements are refined into module partitioning, datapath structure, and control strategy. Here, LLMs can assist with module decomposition, design planning, and clearer expression of design intent.

Following the architecture definition, an initial system model is typically developed using high-level languages like C/C++ or MATLAB, which are used to describe algorithm behavior. A system model is much closer to executable semantics, and its simulation results can therefore serve as a more concrete reference for subsequent hardware implementation. In this stage, LLMs can support the generation of high-level behavioral prototypes \cite{8.8} and support later consistency checking between different representations.

HDL design lies at the center of the front-end flow. As shown in Fig.~\ref{fig:agentic}, there are two primary paths toward an HDL implementation. One path translates the design specification directly into Verilog or VHDL. Although this approach provides fine-grained control over circuit optimization, it requires substantial engineering effort. Moreover, direct LLM-based HDL generation along this path still suffers from limited functional correctness. The other path relies on high-level synthesis (HLS) \cite{24.1} to convert system-level models, such as C/C++ and MATLAB, into HDL designs However, the quality of the generated circuits remains constrained by both the capabilities of the HLS toolchain and the quality of the input HLS code \cite{24.2}. 

Verification is another key stage in front-end design, as it ensures design correctness. As shown in Fig.~\ref{fig:agentic}, testbenches and properties can be constructed from the specification. A testbench provides input stimuli and response checking, while properties capture timing relations and key behavioral constraints. At this stage, LLMs have shown strong promise in testbench generation \cite{14}, failure analysis \cite{16.1}, bug localization \cite{16.2}, and assertion completion \cite{16.0}.

After functional verification, the design proceeds to netlist generation and physical design. Recent studies have also shown that agentic AI can improve automation across backend workflows \cite{33}. As indicated by the consistency-check path in Fig.~\ref{fig:agentic}, the system model also provides a key reference for testing by checking implemented behavior against the original functional intent.

Taken together, these dependencies across specification, modeling, implementation, and verification reveal a central challenge in chip design: the need to preserve semantic consistency across heterogeneous design representations, rather than merely automating isolated tasks. From this perspective, the value of LLMs in EDA lies not only in generating stage-specific artifacts, but also in connecting stages through tool coordination, feedback interpretation, and iterative debugging. With the emergence of agentic AI, their role should evolve beyond localized assistance toward verifiable, traceable, and closed-loop EDA agents.

\vspace{-0.1cm}
\section{LLM for EDA in Front-End Design}\label{sec:1third}

LLM-based front-end design follows a tightly coupled loop of design generation, testing, and repair. This section therefore focuses on key tasks such as the automatic generation of circuit designs and their corresponding testbenches from a shared specification, and further illustrates how LLMs can improve design quality in high-level synthesis. Based on these examples, we evaluate both the practical progress achieved so far and analyze how far current developments advance LLM-assisted front-end EDA.

\vspace{-0.1cm}
\subsection{LLM-based HDL Design Generation}

\begin{figure}[t]
\vspace{0.15cm}
\centering	\includegraphics[width=1\linewidth]{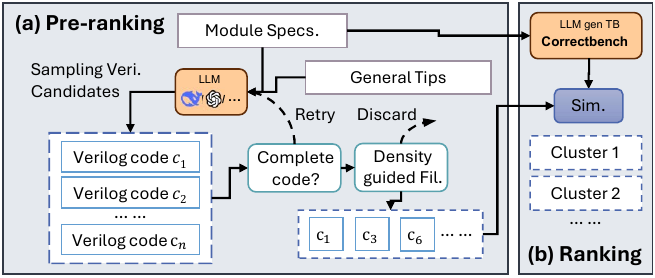}
\vspace{-0.7cm}
	\caption{Overview of Automated HDL Design with Density-Guided Filtering and Simulation-Based Ranking~\cite{10, 11}.}
	\label{fig:vrank}
\vspace{-0.3cm}
\end{figure}

\begin{figure}[t]
\vspace{0cm}
\begin{flushleft}
\includegraphics[width=1\linewidth]{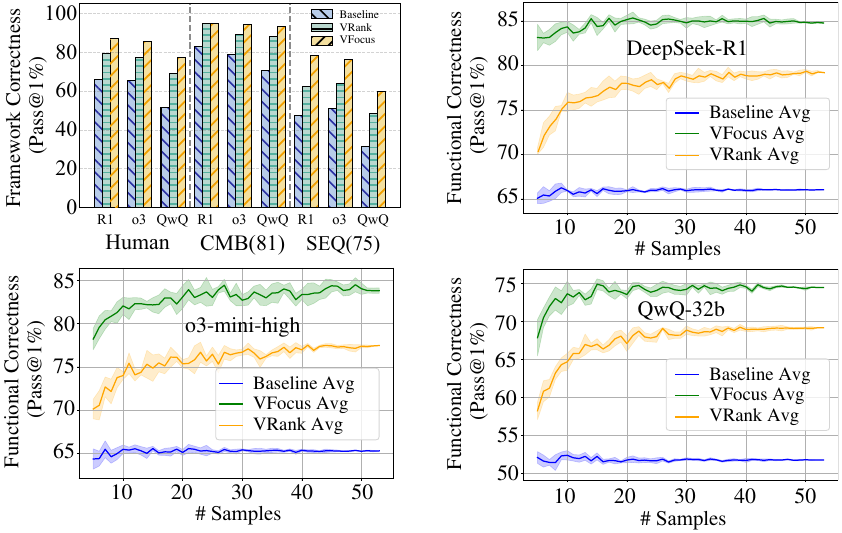}
\end{flushleft}
\vspace{-0.5cm}
\caption{Comparison of Baseline, VRank, and VFocus for HDL Design Generation in Terms of Pass Rate~\cite{11}.}
\label{fig:vfocus}
\vspace{-0.5cm}
\end{figure}

Applying LLMs to HDL generation remains challenging because hardware design requires strict functional correctness, while LLMs are still prone to hallucinations.
To improve the quality of LLM-generated HDL design, VRank \cite{10} proposes a consistency-based framework for Verilog generation. The framework first uses LLMs to generate multiple HDL design candidates, then clusters and ranks them based on their simulation outputs under a shared testbench. Candidates with similar outputs are treated as functionally consistent, and larger clusters are considered more reliable. This allows VRank to more effectively select functionally correct designs. On the VerilogEval benchmark, VRank improves the functional correctness of the generated HDL designs by 10.5\%, demonstrating its effectiveness in improving the accuracy of HDL generation.

Building on this idea, VFocus \cite{11} further shows that ranking candidate code only by simulation results is still not enough. In some cases, the problem is not simply that the model selects the wrong code candidate, but that it fails to reason clearly about key points from the beginning. To address this issue, as shown in Fig. \ref{fig:vrank}, VFocus first analyzes the number of tokens used during the reasoning process for each generated HDL design candidate. The observation suggests that short reasoning often indicates insufficient analysis, while overly long reasoning may drift away from the design intent. Based on this insight, VFocus normalizes the reasoning lengths of different candidates for the same task and keeps only those within a suitable range. It then performs simulation-based clustering and ranking on this refined candidate set. As reported in Fig. \ref{fig:vfocus}, with this pre-ranking refinement, VFocus achieves a 30.9\% improvement over the DeepSeek-R1 baseline and a 16.3\% improvement over the self-consistency method used in VRank.

Taken together, these results indicate an initial breakthrough in localized HDL generation quality under benchmarked and feedback-driven settings. For more complex HDL designs, a more promising future direction is to introduce hierarchical generation and verification mechanisms. In such a framework, candidate implementations for each submodule can first be clustered and ranked, so that functionally more consistent and higher-quality modules are retained. These selected submodules can then be used to construct and optimize the top-level design. Such a strategy may further improve the accuracy of LLM-based generation for large-scale HDL designs.

\vspace{-0.1cm}
\subsection{LLM-based Testbench Generation}
Following the discussion on HDL design generation, another key task in front-end design is testbench construction, which forms the basis of simulation-based hardware verification. Although LLMs have shown great potential in automating HDL design, directly using them to generate testbenches often leads to low pass rates. 

\begin{figure}[t]
\vspace{-0.3cm}
\centering	\includegraphics[width=0.98\linewidth]{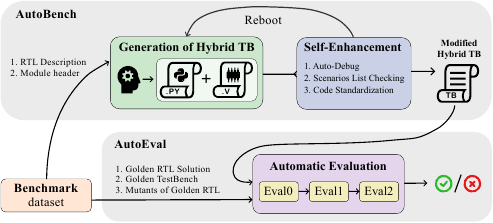}
\vspace{-0.35cm}
	\caption{\fontsize{8.5pt}{9pt}\selectfont Overview of Automated Testbench Generation~\cite{14,15,16}.}
	\label{fig:autobench}
\vspace{-0.35cm}
\end{figure}

\begin{figure}[t]
\vspace{0cm}
\begin{flushleft}
\includegraphics[width=0.98\linewidth]{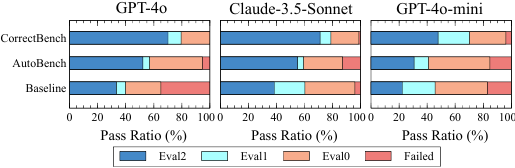}
\end{flushleft}
\vspace{-0.35cm}
\caption{Results of Testbench Generation in Pass Rate~\cite{15}.}
\label{fig:correctbench1}
\vspace{-0.55cm}
\end{figure}

To improve the correctness of LLM-generated testbenches, as shown in Fig. \ref{fig:autobench}, AutoBench \cite{14} improves testbench generation by splitting the task into two parts: a driver and a checker. The driver is responsible for constructing test scenarios, driving the design under test (DUT), and exporting signals, while the checker generates the expected outputs and checks whether the DUT behavior matches them. Notably, the checker is implemented in Python rather than Verilog. This design choice is important for two reasons. First, Python is better suited for expressing high-level checking logic. Second, LLMs are generally more stable in generating Python code than HDLs. AutoBench further introduces AutoEval, which evaluates the quality of the modified hybrid testbench. As shown in Fig. \ref{fig:correctbench1}, compared with the baseline that directly generates the full testbench, AutoBench improves pass@1 by 57\%. On 75 sequential circuit tasks, its performance reaches 3.36× that of the baseline.


CorrectBench \cite{15} further extends testbench generation into a closed-loop self-correction process. It introduces a functional self-validator and an automatic repair module: the framework first generates imperfect HDL designs from the same specification, then uses them to check whether the generated testbench produces decisions consistent with simulation results across different scenarios. If mismatches appear in most HDL designs, the error is more likely to come from the testbench itself. CorrectBench then uses the LLM to revise the problematic parts of the testbench. As reported in Fig. \ref{fig:correctbench1}, CorrectBench achieves a 70.13\% overall pass rate, which is significantly higher than the 52.18\% achieved by AutoBench. 

ConfiBench \cite{16} further improves robustness through scenario masking and an ensemble of multiple generated testbenches, achieving a 72.22\% overall pass rate, higher than CorrectBench’s 70.13\%.

These advances represent meaningful progress in automating a traditionally labor-intensive verification task, especially when generation is combined with decomposition and self-correction. Still, the broader vision of agentic verification remains only partially realized, because coverage closure, corner-case completeness, and robust co-evolution with design generation are not yet solved.

\begin{figure}[t]
\vspace{-0.3cm}
\centering	
\includegraphics[width=0.85\linewidth]{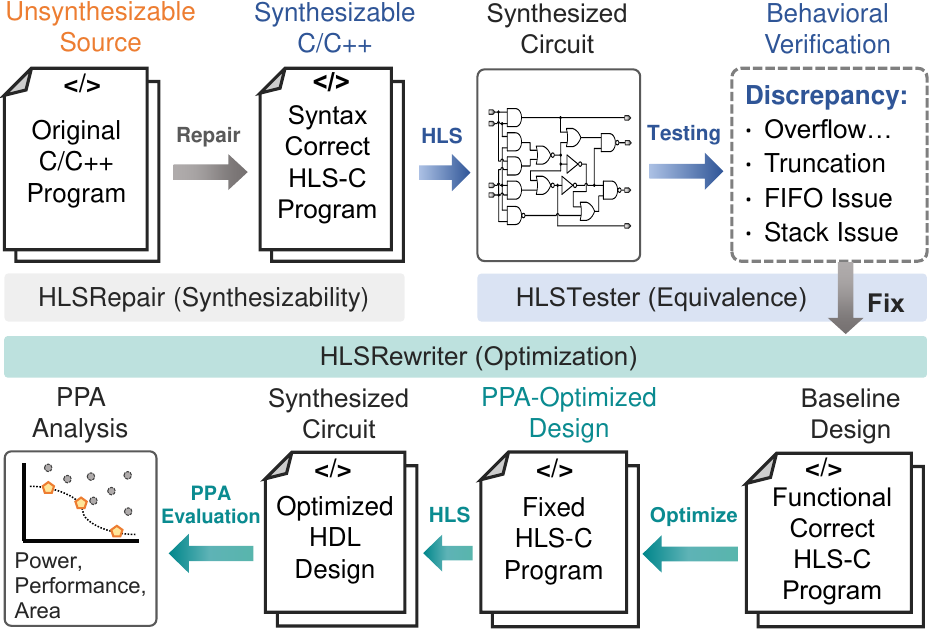}
\vspace{-0.35cm}
	\caption{Overview of an Agentic HLS Flow~\cite{25, 26, 27}.}
	\label{fig:hls}
\vspace{-0.35cm}
\end{figure}

\begin{figure}[t]
\vspace{0cm}
\begin{flushleft}
\includegraphics[width=0.98\linewidth]{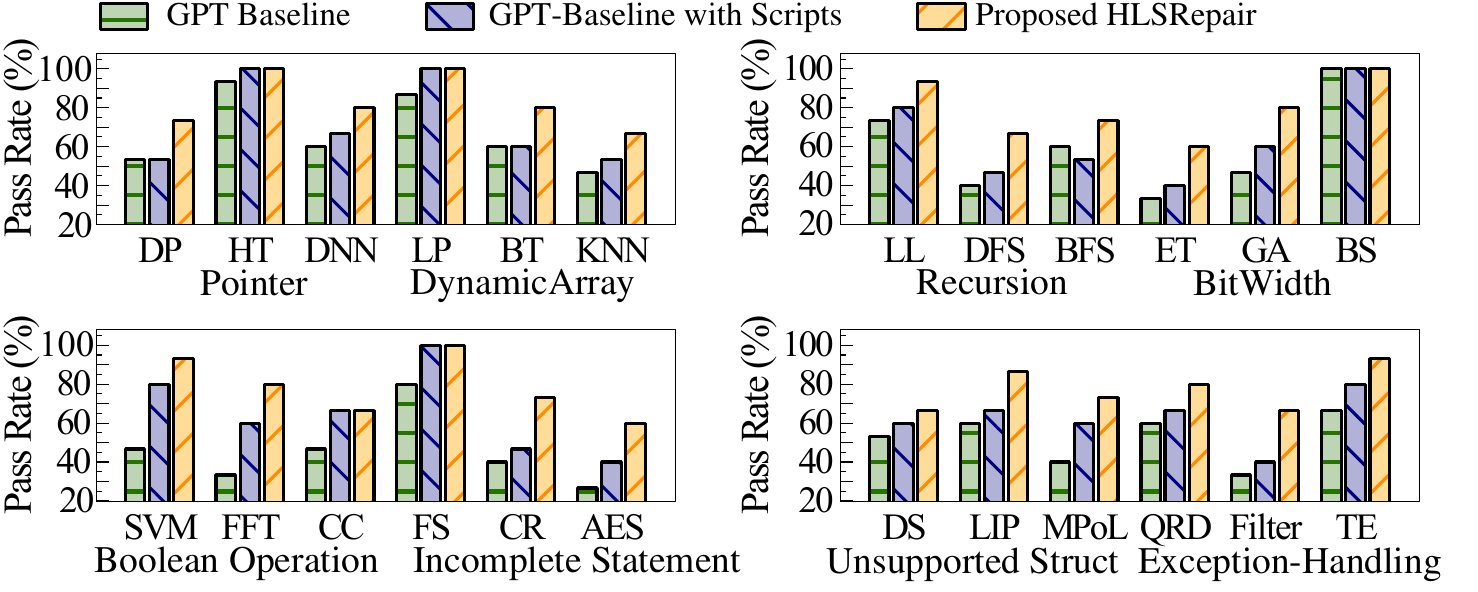}
\end{flushleft}
\vspace{-0.5cm}
\caption{Results of HLS Program Repair in Pass Rate~\cite{25}.}
\label{fig:pass}
\vspace{-0.55cm}
\end{figure}

\vspace{-0.1cm}
\subsection{LLM-based High-Level Synthesis}
High-Level Synthesis (HLS) has emerged as a prominent methodology that translates C/C++ programs into hardware description languages (HDLs), significantly shortening hardware development cycles \cite{24.1}. However, realizing the full potential of HLS is currently hindered by several fundamental barriers, the Compatibility Gap, the Equivalence Gap, and the Optimization Gap \cite{25,26,27}.

As shown in Fig. \ref{fig:hls}, to bridge the Compatibility Gap, HLSRepair introduces an LLM-based repair framework that transforms regular C/C++ programs into synthesizable HLS-C versions \cite{25}. To mitigate LLM hallucinations, a Retrieval-Augmented Generation mechanism is introduced to guide the LLMs toward correct repair. 

To resolve the Equivalence Gap, HLSTester introduces an LLM-assisted testing framework to detect behavioral discrepancies between original C/C++ programs and synthesized hardware \cite{26}. The framework leverages existing C/C++ testbenches to guide the LLM in generating HLS-compatible counterparts. Test inputs are produced via dynamic mutation, with a reasoning chain to increase the likelihood of revealing discrepancies.

To overcome the Optimization Gap, HLSRewriter introduces an LLM-aided program refactoring framework \cite{27}. A decomposition strategy splits complex loop structures into smaller tasks, enabling efficient pipelining. It also integrates a bit-width optimization to balance the trade-off between precision and resource usage, alongside LLM-guided pragma tuning to improve hardware performance.

As reported in Fig.~\ref{fig:pass}, the proposed framework improves the repair pass rate by 23.33\% over the direct use of the LLM. It also achieves 2.71× higher testing efficiency than traditional methods, while improving hardware efficiency, with average reductions of 24.99\%, 12.69\%, and 18.34\% in area, power, and latency, respectively.

Taken together, LLM applications in HLS have shown early tangible progress, benefiting from structured feedback from compilation, synthesis, and QoR evaluation. However, the field is still in its early stage, with challenges remaining in code quality \cite{39}, design optimization, and integration into realistic flows. A promising next step is to develop LLM-enhanced HLS agents that learn from paired HLS–HDL designs and improve through closed-loop feedback \cite{39.1}.

\vspace{-0.1cm}
\section{Challenges and Opportunities}\label{sec:fourth}

Agentic AI has recently emerged as a promising technology for building more autonomous and interactive intelligent systems. Pioneering systems such as OpenClaw point to a promising direction for developing LLM-based systems by organizing model capabilities through retrieved skills and tool-use mechanisms.

\begin{figure}[t]
\vspace{0.1cm}
\centering	\includegraphics[width=1\linewidth]{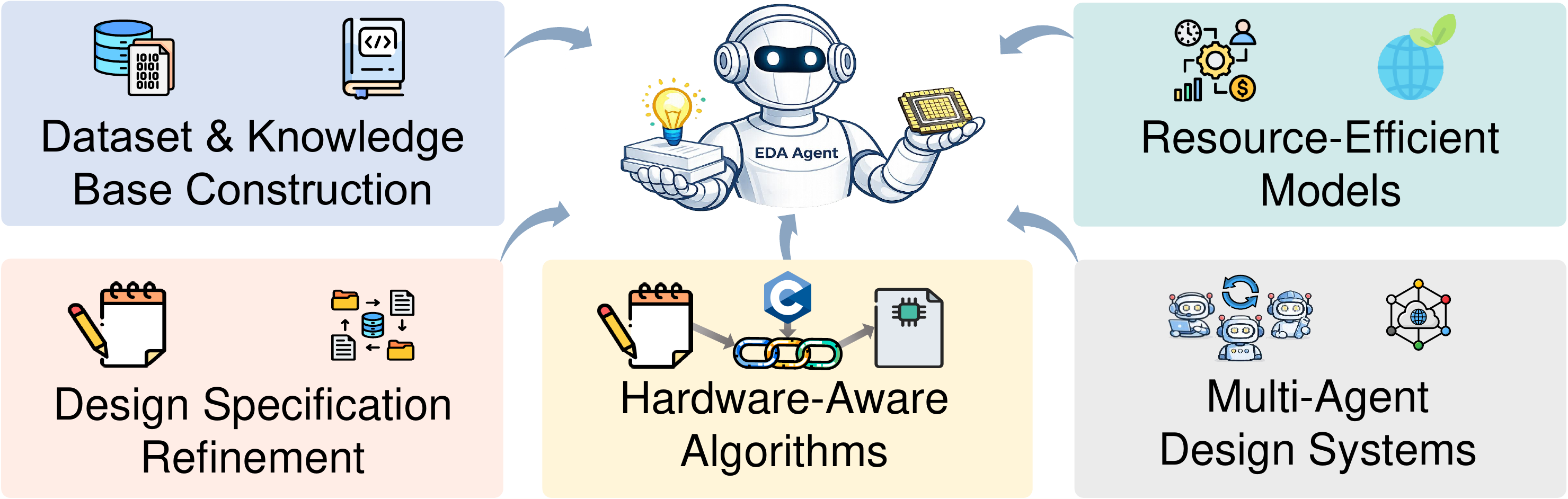}
	\vspace{-0.65cm}
	\caption{~Future Opportunities for EDA in Front-End Design.}
	\label{fig:future}
    \vspace{-0.45cm}
\end{figure}

However, current OpenClaw-style agents still face major limitations when applied to EDA, and several challenges still need to be addressed before such systems can be reliably deployed in real industrial workflows. A major limitation is their difficulty in maintaining semantic consistency across design stages. Another major limitation is their high token consumption and low execution efficiency, since the model often has to process a large volume of lengthy skill documents. To build an OpenClaw-style framework suitable for EDA, the agent should focus on tasks that traditional scripts cannot easily handle, such as specification understanding, repair strategy generation, and cross-stage feedback interpretation, while deterministic scripts and EDA tools continue to handle simulation, synthesis, and data processing. At the same time, long skill descriptions should be compressed into structured and retrievable tool interfaces to reduce context overhead. Persistent state management, design memory, module-level tracing, and verifiable closed-loops are also essential. From this perspective, Fig.~\ref{fig:future} highlights several future directions that can be further explored to advance the development of agentic AI technologies in EDA.

$\circ$~\textit{\textbf{Dataset and Knowledge Base Construction:}} The performance of LLMs in EDA strongly depends on training data, but high-quality hardware datasets remain limited \cite{30.2}. Compared with software repositories, hardware repositories are smaller, more fragmented, and often lack aligned annotations, such as specification-code pairs or verified C++-Verilog mappings. EDA corpora should also include tool manuals, user guides, and report documentation, which provide essential knowledge for tool use and feedback interpretation. Future work should therefore focus on building large-scale, diverse, and well-aligned datasets, together with structured and retrievable knowledge bases, for hardware design tasks.

$\circ$~\textit{\textbf{Design Specification Refinement:}} Writing clear and accurate architecture specifications is a critical first step in chip design and is typically undertaken by experienced engineers. However, ambiguities or minor errors in specifications can still propagate through downstream design steps. Once such issues are exposed at lower levels of abstraction, they become significantly more difficult and costly to diagnose and correct. Therefore, specification refinement is an important direction for future research. LLMs may help improve specification quality through rewriting, clarification, consistency checking, and intent completion, thereby enhancing both the efficiency and reliability of the overall design workflow.

$\circ$~\textit{\textbf{From Natural Language to Hardware-Aware Algorithms:}} Recent work shows that LLMs can translate natural language into C/C++ with high accuracy \cite{38}. A natural next step is to generate hardware-aware C/C++ directly from natural language specifications, thereby improving the accuracy and performance of downstream HDL designs generated through HLS. This direction can support an end-to-end flow from design intent to hardware implementation. Future work should also teach models to learn from expert design strategies, and common HDL patterns, allowing generated designs to better match experienced engineers’ practices.

$\circ$~\textit{\textbf{Collaborative Multi-Agent Design Systems:}} 
Front-end EDA involves tightly connected tasks such as architecture design, HDL generation, verification, and debugging. A single model is often not enough to handle all of them \cite{18.4}. Future research should explore collaborative multi-agent systems, where different agents specialize in different tasks and coordinate toward shared design goals. With clear communication and role division, such systems may better reflect real engineering teams and improve end-to-end automation, especially for complex projects that require cross-stage reasoning.

$\circ$~\textit{\textbf{Resource-Efficient Models for EDA:}} 
As agentic EDA workflows grow in complexity and rely on multiple specialized models and frequent tool interactions, efficiency becomes a major concern. The challenge is not only to reduce model size, but also to jointly optimize model routing and reasoning depth. Future work should explore compact task-specific models, adaptive inference, and cost-aware reasoning strategies, so that LLM-based EDA systems can improve design quality without introducing excessive computational overhead in the practical front-end flow.

\vspace{-0cm}
\section{Conclusion}\label{sec:fifth}
LLMs are opening a new direction for front-end EDA. As discussed in this paper, they have shown strong potential in key tasks such as specification understanding, HDL design generation, testbench construction, and HLS debugging. More importantly, the rise of agentic AI suggests that LLMs may evolve from local assistants into system-level agents that can coordinate tools, interpret feedback, and support closed-loop design flows. Although significant challenges remain, LLMs still hold strong promise for the future of front-end EDA. Realizing this vision will require not only stronger models, but also better datasets, tighter tool integration, and more reliable and verifiable agent frameworks. With continued progress along these directions, LLM-enabled EDA could become an important step toward more automated and intelligent chip design.

 

\newpage

\begin{thebibliography}{9}
\balance


\bibitem{8.4} M. Li, W. Fang, Q. Zhang, and Z. Xie, ``SpecLLM: Exploring Generation and Review of VLSI Design Specification with Large Language Model,'' in \textit{Proceedings of the International Symposium of Electronics Design Automation (ISEDA)}, 2025.
\bibitem{8.70} M. Liu \textit{et al.}, ``ChipNeMo: Domain-Adapted LLMs for Chip Design,'' \textit{arXiv preprint arXiv:2311.00176}, 2023.
\bibitem{4} M. Liu, N. R. Pinckney, B. Khailany, and H. Ren, ``Invited Paper: VerilogEval: Evaluating Large Language Models for Verilog Code Generation,'' in \textit{Proceedings of the IEEE/ACM International Conference on Computer-Aided Design (ICCAD)}, 2023.
\bibitem{13.1} Z. Zhang, B. Szekely, P. Gimenes, G. Chadwick, H. McNally, J. Cheng, R. D. Mullins, and Y. Zhao, ``LLM4DV: Using Large Language Models for Hardware Test Stimuli Generation,'' in \textit{Proceedings of the IEEE International Symposium on Field-Programmable Custom Computing Machines (FCCM)}, 2025.

\bibitem{16.00} H. Wu, Z. He, X. Zhang, X. Yao, S. Zheng, H. Zheng, and B. Yu, ``ChatEDA: A Large Language Model Powered Autonomous Agent for EDA,'' \textit{IEEE Transactions on Computer-Aided Design of Integrated Circuits and Systems}, vol. 43, no. 10, 2024.
\bibitem{16.1} S. Qiu, M. Wang, R. Afsharmazayejani, M. M. Shahmiri, B. Tan, H. Pearce, ``Towards LLM-based Root Cause Analysis of Hardware Design Failures,'' in \textit{Proceedings of the IEEE International Conference on Omni-layer Intelligent Systems}, 2025.
\bibitem{16.2} J. Li, S. -Z. Wong, G. -W. Wan, X. Wang and J. Yang, ``EDA-Debugger: An LLM-Based Framework for Automated EDA Runtime Issue Resolution,'' in \textit{Proceedings of the IEEE International Symposium on Quality Electronic Design (ISQED)}, 2025.
\bibitem{18.3} OpenClaw, ``OpenClaw,'' GitHub repository, 2026. [Online]. Available: https://github.com/openclaw/openclaw
. Accessed: Apr. 15, 2026.
\bibitem{18.4} K. Xu, J. Sun, Y. Hu, X. Fang, W. Shan, X. Wang, and Z. Jiang, ``MEIC: Re-thinking RTL Debug Automation using LLMs,'' in \textit{Proceedings of the IEEE/ACM International Conference on Computer-Aided Design (ICCAD)}, 2024.
\bibitem{b18.5} K. Xu, R. Qiu, Z. Zhao, G. L. Zhang, U. Schlichtmann, B. Li, “LLM-Aided Efficient Hardware Design Automation,” \textit{arXiv: 2410.18582}, 2024.
\bibitem{14} R. Qiu, G. L. Zhang, R. Drechsler, U. Schlichtmann, and B. Li, ``AutoBench: Automatic Testbench Generation and Evaluation Using LLMs for HDL Design,'' in \textit{Proceedings of the ACM/IEEE International Symposium on Machine Learning for CAD (MLCAD)}, 2024.
\bibitem{15} R. Qiu, G. L. Zhang, R. Drechsler, U. Schlichtmann, and B. Li, ``CorrectBench: Automatic Testbench Generation with Functional Self-Correction using LLMs for HDL Design,'' in \textit{Proceedings of the IEEE/ACM Design, Automation \& Test in Europe Conference \& Exhibition (DATE)}, 2025.
\bibitem{16} R. Qiu, G. L. Zhang, R. Drechsler, T.-Y. Ho, U. Schlichtmann, and B. Li, ``ConfiBench: Automatic Testbench Generation with Confidence-Based Scenario Mask and Testbench Ensemble using LLMs for HDL Design,'' \textit{ACM Transactions on Design Automation of Electronic Systems (TODAES)}, 2026.


\bibitem{8.8} J. Gai, H. M. Chen, Z. Wang, H. Zhou, W. Zhao, N. Lane, and H. Fan, ``Exploring Code Language Models for Automated HLS-based Hardware Generation: Benchmark, Infrastructure and Analysis,'' in \textit{Proceedings of the IEEE/ACM Asia and South Pacific Design Automation Conference (ASP-DAC)}, 2025.
\bibitem{24.1} J. Cong, J. Lau, G. Liu, S. Neuendorffer, P. Pan, K. Vissers, and Z. Zhang, ``FPGA HLS Today: Successes, Challenges, and Opportunities,'' \textit{ACM Transactions on Reconfigurable Technology and Systems (TRETS)}, vol. 15, no. 4, 2022.
\bibitem{24.2} S. Lahti, P. Sjövall, J. Vanne, T. Hämäläinen, ``Are We There Yet? A Study on the State of High-Level Synthesis,'' \textit{IEEE Transactions on Computer-Aided Design of Integrated Circuits and Systems (TCAD)}, 2019.
\bibitem{16.0} Z. Yan, W. Fang, M. Li, M. Li, S. Liu, Z. Xie, H. Zhang, ``AssertLLM: Generating Hardware Verification Assertions from Design Specifications via Multi-LLMs,'' in \textit{Proceedings of the IEEE/ACM Asia and South Pacific Design Automation Conference (ASP-DAC)}, 2025.
\bibitem{33} A. Ghose, A. B. Kahng, S. Kundu, B. Pramanik, ``Invited: Agentic AI for Physical Design R\&D: Status and Prospects,'' in \textit{Proceedings of the IEEE International Symposium on Physical Design (ISPD)}, 2026.
\bibitem{10} Z. Zhao, R. Qiu, I.-C. Lin, G. L. Zhang, B. Li, U. Schlichtmann, ``VRank: Enhancing Verilog Code Generation from LLMs via Self-Consistency,'' in \textit{Proceedings of the IEEE International Symposium on Quality Electronic Design (ISQED)}, 2025.
\bibitem{11} Z. Zhao, B. Li, G. L. Zhang, and U. Schlichtmann, ``VFocus: Better Verilog Generation from Large Language Model via Focused Reasoning,'' in \textit{Proceedings of the IEEE International System-on-Chip Conference (SOCC)}, 2025.

\bibitem{25} K. Xu, G. L. Zhang, X. Yin, C. Zhuo, U. Schlichtmann, and B. Li, ``Automated C/C++ Program Repair for High-Level Synthesis via Large Language Models,'' in \textit{Proceedings of the ACM/IEEE International Symposium on Machine Learning for CAD (MLCAD)}, 2024.
\bibitem{26} K. Xu, B. Li, G. L. Zhang, and U. Schlichtmann, ``HLSTester: Efficient Testing of Behavioral Discrepancies with LLMs for High-Level Synthesis,'' in \textit{Proceedings of the IEEE/ACM International Conference on Computer-Aided Design (ICCAD)}, 2025.
\bibitem{27} K. Xu, G. L. Zhang, X. Yin, C. Zhuo, U. Schlichtmann, and B. Li, ``HLSRewriter: Efficient Refactoring and Optimization of C/C++ Code with LLMs for High-Level Synthesis,'' \textit{ACM Transactions on Design Automation of Electronic Systems}, 2026.
\bibitem{27.1} K. Xu, et al., “Logic Design of Neural Networks for High-Throughput and Low-Power Applications,'' in \textit{Proceedings of the IEEE/ACM Asia and South Pacific Design Automation Conference (ASP-DAC)}, 2024.
\bibitem{30.2} Q. Xu, L. Stok, R. Drechsler, X. Wang, G. L. Zhang, I. L. Markov, ``Revolution or Hype? Seeking the Limits of Large Models in Hardware Design,'' in \textit{Proceedings of the IEEE/ACM International Conference on Computer-Aided Design (ICCAD)}, 2025.

\bibitem{30.2} Q. Xu, L. Stok, R. Drechsler, X. Wang, G. L. Zhang, I. L. Markov, ``Revolution or Hype? Seeking the Limits of Large Models in Hardware Design,'' in \textit{Proceedings of the IEEE/ACM International Conference on Computer-Aided Design (ICCAD)}, 2025.
\bibitem{38} Z. Yu, Y. Zhao, A. Cohan, and X.-P. Zhang, ``HumanEval Pro and MBPP Pro: Evaluating Large Language Models on Self-invoking Code Generation Task,'' in \textit{Findings of the Association for Computational Linguistics (ACL)}, 2025.
\bibitem{39} K. Xu, G. L. Zhang, U. Schlichtmann, and B. Li, “CorrectHDL: Agentic HDL Design with LLMs Leveraging High-Level Synthesis as Reference,” \textit{arXiv preprint arXiv:2511.16395}, 2025.
\bibitem{39.1} K. Xu, D. Schwachhofer, J. Blocklove, I. Polian, P. Domanski, D. Pfluger, S. Garg, R. Karri, O. Sinanoglu, J. Knechtel, Z. Zhao, U. Schlichtmann and B. Li, ``Large Language Models (LLMs) for Electronic Design Automation (EDA) : Special Session Paper,'' in \textit{Proceedings of the IEEE International System-on-Chip Conference}, 2025.








































\end{thebibliography}
\end{document}